%% file: Herleitung.tex
\documentclass[a4paper,   DIV=15,BCOR=8.25mm, fontsize=12pt, headings=openright, bibliography=totocnumbered, twoside=off,  headsepline,  numbers=noenddot,  enabledeprecatedfontcommands]{scrartcl}
\pdfoutput=1
\usepackage[utf8]{inputenc} 
\usepackage[T1]{fontenc}    
\usepackage[english]{babel} 
\usepackage{bibgerm}

\usepackage{scrtime}

\usepackage{enumitem}

\makeatletter
\g@addto@macro{\@afterheading}{\vspace{-\parskip}}
\makeatother

\usepackage[automark,headsepline]{scrpage2}
\clearscrheadings
\clearscrplain
\ohead{\headmark}
\ofoot[\pagemark]{\pagemark}
\setheadsepline{.4pt} 
\usepackage[marginal]{footmisc}

\usepackage{textcomp}           
\usepackage{amsfonts}
\usepackage{amssymb}            
\usepackage{dsfont}
\usepackage[printonlyused]{acronym}
\usepackage{wrapfig} 
\usepackage{amsmath} 
\usepackage{upgreek}            
\usepackage{ziffer}             
\usepackage{cancel}
\usepackage{ulem}

\usepackage[pdftex]{graphicx} 
\usepackage{rotating}	      
\usepackage{xspace}

\usepackage{color,colortbl}
\definecolor{green4}{rgb}{0, 0.54, 0}
\definecolor{Green}{rgb}{0.8, 1.0, 0.8}
\definecolor{Red}{rgb}{1.0, 0.8, 0.8}
\definecolor{Gray}{gray}{0.9}
\definecolor{Gray2}{gray}{0.5}
\definecolor{ggplotRed}{rgb}{0.89, 0.10, 0.11}
\definecolor{ggplotGreen}{rgb}{0.30, 0.69, 0.29}
\definecolor{ggplotBlue}{rgb}{0.22, 0.49, 0.72}

\usepackage{footnote}
\makesavenoteenv{tabular}
\makesavenoteenv{table}

\usepackage{bookmark} 
\usepackage{pdfpages}
\usepackage[table]{xcolor}
\usepackage{colortbl}
\usepackage{multirow}
\usepackage{booktabs}
\usepackage{capt-of}
\usepackage{float}     
\usepackage{subfig}    
\usepackage[noadjust]{cite}

\usepackage{longtable} 
\usepackage{tabularx} 
\newcolumntype{L}[1]{>{\raggedright\arraybackslash}p{#1}} 
\newcolumntype{C}[1]{>{\centering\arraybackslash}p{#1}} 
\newcolumntype{R}[1]{>{\raggedleft\arraybackslash}p{#1}} 
\usepackage{placeins} 

\usepackage{siunitx}
\usepackage{chngcntr} 

\usepackage[NoDate]{currvita}
\AtBeginDocument{%
	\settowidth{\cvlabelwidth}{%
		\cvlabelfont Programmiersprachen:%
	}%
}%

\renewcommand{\j}{\mathrm{j}}                       
\newcommand{\e}{\mathrm{e}}                       

\DeclareFontFamily{U}{wncy}{}
\DeclareFontShape{U}{wncy}{m}{n}{<->wncyr10}{}
\DeclareSymbolFont{mcy}{U}{wncy}{m}{n}
\DeclareMathSymbol{\Sha}{\mathord}{mcy}{"58} 

\usepackage{listings}
\usepackage{color} 
\definecolor{mygreen}{RGB}{28,172,0} 
\definecolor{mylilas}{RGB}{170,55,241}

\makeatletter
\renewcommand\paragraph{\@startsection{paragraph}{4}{\z@}%
	{-3.25ex\@plus -1ex \@minus -.2ex}%
	{1.5ex \@plus .2ex}%
	{\normalfont\normalsize\bfseries}}
\makeatother

\let\tempone\itemize
\let\temptwo\enditemize
\renewenvironment{itemize}{\tempone\addtolength{\itemsep}{-0.5\baselineskip}}{\temptwo}

\title{Derivation of an aggregated band pseudo phasor for single phase pulse width modulation voltage waveforms}
\author{Matthias Klatt, Jan Meyer and Peter Schegner}
\date{}

\begin{document}
\maketitle

\input{inhalt.tex}

\bibliographystyle{gerunsrt} \bibliography{Buecher_Diss} 	
\end{document}

%% file: inhalt.tex
\section{Motivation}

Holmes' and Lipo's book \cite{BookHolmes2003} on pulse width modulation (PWM) is a comprehensive work regarding the use of PWM in power converters. Among many other aspects, it contains the mathematical derivation of voltage waveform for PWM with different kinds of sampling, carrier shapes, number of phases, voltage and current source inverters, or even for multilevel converters.

All equations of the voltage waveforms in the book are given in a way so they directly represent their spectral components. However, for the purpose of supraharmonic modeling under certain conditions it may be more suitable to see these individual components as integral parts of the corresponding emission bands. The goal of this paper is to change the original equation to show that all components in an emission band share a common frequency and phase relation. This proves that each emission band can be seen as a single spectral component with a well defined RMS value, frequency and phase, but which is amplitude-modulated within a fundamental period of the network frequency. 
This fact can then be used as the mathematical basis to describe each emission band as a single, aggregated pseudo-phasor.
This result is necessary as the mathematical substantiation for simplifications in future supraharmonic modeling approaches. 

Phasors in general are defined to have a constant magnitude. The newly defined pseudo-phasor, however, is amplitude modulated but with a constant RMS value over interger periods of the network frequency. The pseudo-phasor is therefore no longer compatible with the definition of the frequency domain or the Fourier Transform, but is still valid within the Hilbert domain.

\pagebreak
\begin{figure}[b]
	\centering
	\includegraphics[width=8cm]{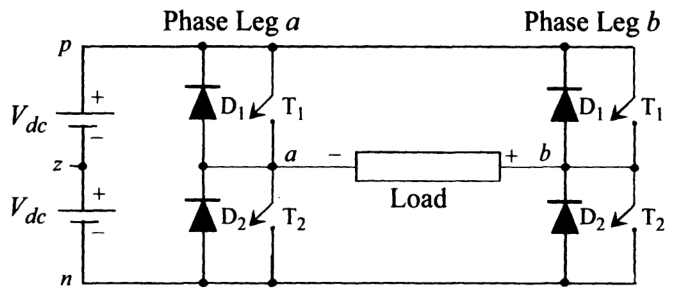}
	\caption{Equivalent circuit of a power converter H-bride with split DC link voltage,\newline from \cite{BookHolmes2003}, Figure 4.1 on page 156}
	\label{fig:ESB_single_leg}
\end{figure}
\section{Derivation}
The voltage controller output (CO), which is the intended voltage at the H-bridge terminals a and b (see Figure~\ref{fig:ESB_single_leg}) shall be defined as
\begin{align}
u_\mathrm{ab\,CO}(t) = U_\mathrm{ab\,CO}\cdot \sqrt{2}\,\cos(\omega_\mathrm{0}\,t+\varphi_\mathrm{0}).
\end{align}
The mathematical represenation of the waveform of the voltage of single-phase voltage-source inverters using PWM with natural sampling and a triangular carrier is given in \cite{BookHolmes2003} in equation (4.6) on page 160 as
\begin{align}
	u_\mathrm{ab} (t) =	&~ 2 \cdot U_\mathrm{dc} \cdot  M \cdot \cos\left(\omega_0 \cdot t\right)+ \nonumber \\
						&~ \frac{8\,U_\mathrm{dc}}{\pi} \sum_{m=1}^{\infty} \sum_{n=-\infty}^{\infty} \frac{1}{2\,m} \mathrm{J}_{2\,n-1}(m\,\pi\,M) \cdot \cos\left(\left[m+n-1\right]\pi\right) \cdot \label{eqn:OrgHolmes}\\
						&~ \qquad \qquad \qquad \qquad \qquad \cos \left(2\,m\,\omega_\mathrm{c} \cdot t + \left[2\,n-1\right]\omega_0\cdot t\right)  && m,n\in \mathds{Z} \nonumber
\end{align}
with the modulation index $M$
\begin{align}
	M &= \frac{U_\mathrm{ab\,CO}}{\sqrt{2} \cdot  U_\mathrm{dc}}.
\end{align}
The equation (\ref{eqn:OrgHolmes}) is derived by combining the spectra of two individual converter legs, each using half of the DC link voltage $U_\mathrm{dc}$ (see Figure~\ref{fig:ESB_single_leg}). $\omega_0$ is the AC network frequency and $\omega_\mathrm{c}$ is the carrier signal frequency, and $\mathrm{J}_\chi(x)$ are Bessel functions of the first kind of order~$\chi$.

The following changes to this representation are made:
\begin{itemize}
	\item In this paper the full DC link voltage $U_\mathrm{DC}$ is used.
	\begin{align}
		U_\mathrm{DC} = 2\cdot U_\mathrm{dc}
	\end{align}
	\item The network frequency will be represented as $\omega_\mathrm{N}$.
	\item In \cite{BookHolmes2003}, the phase angles of the carrier and fundamental network voltage are neglected. These are reintroduced as $\varphi_\mathrm{c}$ and $\varphi_\mathrm{N}$.
	\item For integer values of m and n the term $\cos\left(\left[m+n-1\right]\pi\right)$ results in values of one with alternating signs. It is replaced by its equal term 
	\begin{align}
		\cos\left(\left[m+n-1\right]\pi\right) &= {(-1)^{m+n-1}}= \frac{1}{(-1)^{m+n-1}}.
	\end{align}
\end{itemize}
The resulting representation of the voltage at the H-bridge is
\begin{align}
	u_\mathrm{ab} (t) =	&~ U_\mathrm{DC} \cdot  M \cdot \cos\left(\omega_\mathrm{N} \cdot t + \varphi_\mathrm{N}\right)+ \nonumber \\
	& \frac{2\,U_\mathrm{DC}}{\pi} \sum_{m=1}^{\infty} \sum_{n=-\infty}^{\infty} \frac{\mathrm{J}_{2\,n-1}(m\, \pi\, M)}{m \cdot (-1)^{m+n-1}} \cdot \label{eqn:AusgangsgleichungInMeinerDarstellung}\\
	& ~\qquad\qquad\qquad\qquad \cos \left(2\,m \left[\omega_\mathrm{c} \cdot t + \varphi_\mathrm{c}\right] + \left[2\,n-1\right] \left[\omega_\mathrm{N}\cdot t + \varphi_\mathrm{N}\right]\right) \nonumber 
\end{align}
with
\begin{align}
	M &= \frac{\sqrt{2} \cdot U_\mathrm{ab\,CO}}{ U_\mathrm{DC}}.
\end{align}
This can be split into the intended fundamental component $u_\mathrm{ab\,CO}(t)$ and the parasitic supraharmonic distortion (SHD) components $u_\mathrm{ab\,SHD}(t)$.
\begin{align}
	u_\mathrm{ab} (t) =&~u_\mathrm{ab\,CO}(t) + u_\mathrm{ab\,SHD}(t) \\
	u_\mathrm{ab\,CO} (t) =&~U_\mathrm{DC} \cdot  M \cdot \cos\left(\omega_\mathrm{N} \cdot t + \varphi_\mathrm{N}\right) \\
	u_\mathrm{ab\,SHD}(t) =&~\frac{2\,U_\mathrm{DC}}{\pi} \sum_{m=1}^{\infty} \sum_{n=-\infty}^{\infty} \frac{\mathrm{J}_{2\,n-1}(m\,\pi\,M)}{m \cdot (-1)^{m+n-1}} \cdot \nonumber \\
	& ~~\qquad \qquad \qquad \quad \cos \left(2\,m \left[\omega_\mathrm{c} \cdot t + \varphi_\mathrm{c}\right] + \left[2\,n-1\right] \left[\omega_\mathrm{N}\cdot t + \varphi_\mathrm{N}\right]\right) 
	\label{eqn:UmgeformterSHD}
\end{align}
The equation (\ref{eqn:UmgeformterSHD}) shows that the supraharmonic spectrum consists of an infinite number of emission bands of order $m$, each composed of an infinite number of spectral components of order $n$.
For the desired changes in the equation, only the second sum of the supharmarmonic distortion component $\Psi(m,t)$ needs to be considered.
\begin{align}
	u_\mathrm{ab\,SHD}(t) =& ~\frac{2\,U_\mathrm{DC}}{\pi} \sum_{m=1}^{\infty} \Psi(m,t) \label{eqn:Uab_SHD_Psi} \\
	\Psi(m,t) =& \sum_{n=-\infty}^{\infty} \frac{\mathrm{J}_{2\,n-1}(m\,\pi\,M)}{m \cdot (-1)^{m+n-1}} \cdot \cos \left(2\,m \left[\omega_\mathrm{c} \cdot t + \varphi_\mathrm{c}\right] + \left[2\,n-1\right] \left[\omega_\mathrm{N}\cdot t + \varphi_\mathrm{N}\right]\right) \label{eqn:Ausgangsgleichung_SHD}
\end{align}
To shorten the representation, the variables $\alpha$ and $\beta$ are introduced.
\begin{align}
	\alpha(m,t) 	&= 2\,m\left[\omega_\mathrm{c}\cdot t + \varphi_\mathrm{c}\right] \\
	\beta(t)  	&= \left[\omega_\mathrm{N} \cdot t + \varphi_\mathrm{N}\right]
\end{align}
The equation (\ref{eqn:Ausgangsgleichung_SHD}) is simplified and the sum is split at $n\ge1$.
\begin{align}
	\Psi(m,t) 	=& \sum_{n=1}^{\infty} \frac{\mathrm{J}_{2\,n - 1}(m\,\pi\,M)}{m \cdot (-1)^{m+n-1}} \cdot \cos \left(\alpha + \left[2\,n - 1\right] \beta\right) + \nonumber \\
				 & \sum_{n=0}^{-\infty} \frac{\mathrm{J}_{2\,n - 1}(m\,\pi\,M)}{m \cdot (-1)^{m+n-1}} \cdot \cos \left(\alpha + \left[2\,n - 1\right] \beta\right) \label{eqn:AufgeteilteWerte}
\end{align}
The sum for negative values of $n$ in equation (\ref{eqn:AufgeteilteWerte}) is converted, so its index will use the identical values as that for the sum of the positive values using the transformation
\begin{align}
	   n &= -n'+1 \qquad n' \in \mathds{Z} \\
	2\,n-1 &= -2\,n'+1
\end{align}
leading to
\begin{align}
	\Psi(m,t) 	=& \sum_{n=1}^{\infty} \frac{\mathrm{J}_{2\,n - 1}(m\, \pi\, M)}{m \cdot (-1)^{m+n-1}}\cos \left(\alpha + \left[2\,n - 1\right] \beta\right) \nonumber \\ 
				+& \sum_{n'=1}^{\infty} \frac{\mathrm{J}_{-2\,n'+1}(m\, \pi\, M)}{m \cdot (-1)^{m+(-n'+1)-1}} \cos \left(\alpha + \left[-2\,n' + 1\right] \beta\right). \label{eqn:SummemitNStrich}
\end{align}
For integer values of the order $\chi$, Bessel functions of the first kind $\mathrm{J}_\chi(x)$ with negative orders can be converted to positive orders using the relationship
\begin{align}
	\mathrm{J}_{-\chi}(x) &= (-1)^\chi \cdot \mathrm{J}_\chi (x). && \chi \in \mathds{Z}
\end{align}
As all orders of the Bessel functions in the second sum in equation~(\ref{eqn:SummemitNStrich}) $\chi=-2\,n'+1$ are odd, this results in
\begin{align}
	\mathrm{J}_{-2\,n'+1} &= (-1)\cdot \mathrm{J}_{2\,n'-1}\,.
\end{align}
For any values of the integer $\mu$, the relationship
\begin{align}
	(-1)^{\mu} = (-1)^{-\mu} && \mu \in \mathds{Z}
\end{align}
is valid, so the term $(-1)^{m+(-n'+1)-1}$ can be transformed to
\begin{align}
	(-1)^{m+(-n'+1)-1} = (-1)\cdot (-1)^{m+n'-1}\,.
\end{align}
Therefore, the equation for $\Psi(m,t)$ can be changed to the following form
\begin{align}
	\Psi(m,t) 	=& \sum_{n=1}^{\infty} \frac{\mathrm{J}_{2\,n - 1}(m\,\pi\,M)}{m \cdot (-1)^{m+n-1}}\cos \left(\alpha + \left[2\,n - 1\right] \beta\right) \nonumber \\ 
				+& \sum_{n'=1}^{\infty}  \frac{(-1)}{(-1)} \cdot \frac{\mathrm{J}_{2\,n'-1}(m\,\pi\,M)}{m \cdot (-1)^{m+n'-1}} \cos \left(\alpha + \left[-2\,n' + 1\right] \beta\right), 
\end{align}
where the two $(-1)$ cancel out.
Due to the applied substitutions, the sums can now be combined again into a single sum.
\begin{align}
	\Psi(m,t) 	=& \sum_{k=1}^{\infty} \frac{\mathrm{J}_{2\,k - 1}(m\,\pi\,M)}{m \cdot (-1)^{m+k-1}} \cdot \left[\cos \left(\alpha + \left[2\,k - 1\right] \beta\right) + \cos \left(\alpha - \left[2\,k - 1\right] \beta\right) \right] \label{eqn:PsimitCos}
\end{align}
With the trigonometric identity
\begin{align}
	\cos(\delta\pm \gamma) = \cos(\delta)\cos(\gamma)\mp \sin(\delta)\sin(\gamma)
\end{align}
the two cosine terms can be rearranged to
\begin{align}
	\cos \left(\alpha + \left[2\,k - 1\right] \beta\right) + \cos \left(\alpha - \left[2\,k - 1\right] \beta\right) = \qquad \qquad \qquad \qquad \qquad \qquad \qquad \nonumber\\
	\cos(\alpha)\cos(\left[2\,k - 1\right] \beta)-\sin(\alpha)\sin(\left[2\,k - 1\right] \beta) \\
	+ \cos(\alpha)\cos(\left[2\,k - 1\right] \beta) + \sin(\alpha)\sin(\left[2\,k - 1\right] \beta)\,, \nonumber
\end{align}
leading to
\begin{align}
	\cos \left(\alpha + \left[2\,k - 1\right] \beta\right) + \cos \left(\alpha - \left[2\,k - 1\right] \beta\right) = \sqrt{2} \cos(\alpha)\cdot \sqrt{2}\cos(\left[2\,k - 1\right] \beta).\hspace{.25cm} \label{eqn:cosUmformung}
\end{align}
Then the equation (\ref{eqn:cosUmformung}) is interted into (\ref{eqn:PsimitCos}).
\begin{align}
	\Psi(m,t) 	=&  \sqrt{2} \cos(\alpha) \cdot \sum_{k=1}^{\infty} \frac{\mathrm{J}_{2\,k - 1}(m\,\pi\,M)}{m \cdot (-1)^{m+k-1}} \cdot \sqrt{2}\cos(\left[2\,k - 1\right] \beta) \label{eqn:PsiUmgeformtZumEinsetzen}
\end{align}
By inserting equation (\ref{eqn:PsiUmgeformtZumEinsetzen}) into (\ref{eqn:Uab_SHD_Psi}), as well as reversing the substitution of $\alpha$ and $\beta$, the new representation of the supraharmonic distortion component $u_\mathrm{ab\,SHD}(t)$ is found.
\begin{align}
	u_\mathrm{ab\,SHD}(t) = \underbrace{\frac{2\,U_\mathrm{DC}}{\pi}}_\mathrm{(I)}   \cdot \sum_{m=1}^{\infty} & \underbrace{\frac{1}{m}}_\mathrm{(II)}\cdot \underbrace{\sqrt{2}\cos(2\,m\left[\omega_\mathrm{c}\cdot t + \varphi_\mathrm{c}\right])}_\mathrm{(III)}\cdot \nonumber \\
	&\sum_{k=1}^{\infty} \underbrace{\frac{\mathrm{J}_{2\,k - 1}(m\,\pi\,M)}{(-1)^{m+k-1}}}_\mathrm{(IV)} \cdot \underbrace{\sqrt{2}\cos(\left[2\,k - 1\right] \left[\omega_\mathrm{N} \cdot t + \varphi_\mathrm{N}\right])}_\mathrm{(V)} \label{eqn:Ergebnis}
\end{align}
The equation (\ref{eqn:Ergebnis}) consists of five main components, which are interpreted as follows:
\begin{enumerate}[label=(\Roman*)]
	\item Scaling factor: It is dependent of the DC link voltage only.
	\item Emission band scaling: The magnitude of each emission band decreases with increasing order.
	\item Emission band modulation: The emission band of order $m$ has the frequency $2\,m\,\omega_\mathrm{c}$ and the phase $2\,m\,\varphi_\mathrm{c}$. This modulation is neutral regarding the RMS value of the emission band.
	\item Emission band harmonic scaling: Each emission band is formed by an infinite number of harmonics, which are each a combination of a pair of spectral components from equation~(\ref{eqn:AusgangsgleichungInMeinerDarstellung}). This section scales the magnitude of each such harmonic.
	\item Emission band harmonic modulation: This section performs the modulation of each emission band harmonic. It contains only odd harmonics of orders $\nu=\left[2\,k - 1\right]$. The frequency and phase of each harmonic are $\nu\cdot \omega_\mathrm{N}$ and $\nu\cdot \varphi_\mathrm{N}$. This modulation is neutral regarding the RMS value of the emission band harmonic.
\end{enumerate}
\section{Conclusions and Interpretations}
\paragraph{Pseudo-phasor}
The amplitude-modulated, rotating pseudo-phasor $\underline{\tilde{u}}_\mathrm{ab}^{\mathrm{B}m}(t)$ for the emission band of order $m$ is the analytical signal 
\begin{align}
	\underline{\tilde{u}}_\mathrm{ab}^{\mathrm{B}m}(t) = u_\mathrm{ab\,SHD}(m,t) + \j \cdot \mathcal{H}\left\{u_\mathrm{ab\,SHD}(t)\right\}
\end{align}
of the voltage in equation (\ref{eqn:Ergebnis}), where $\mathcal{H}\left\{\chi\right\}$ is the Hilbert transform of $\chi$.
\begin{align}
\underline{\tilde{u}}_\mathrm{ab}^{\mathrm{B}m}(t) = \left[\frac{4\,U_\mathrm{DC}}{m\,\pi}\cdot \sum_{k=1}^{\infty} \frac{\mathrm{J}_{2\,k - 1}(m\,\pi\,M)}{(-1)^{m+k-1}} \cdot \cos(\left[2\,k - 1\right] \left[\omega_\mathrm{N} \cdot t + \varphi_\mathrm{N}\right]) \right] \cdot \e^{\,\j\,2\,m\left[\omega_\mathrm{c}\cdot t + \varphi_\mathrm{c}\right]}
\end{align}
\paragraph{RMS value}
The RMS voltage of the $m^\mathrm{th}$ emission band at the H-bridge terminals $U_\mathrm{ab}^{\mathrm{B}m}$ is the product of the RMS voltages of the sections (I) and (II), and the RMS value of the sum of section~(IV).
\begin{align}
	U_\mathrm{ab}^{\mathrm{B}m} = \frac{2\,U_\mathrm{DC}}{\pi} \cdot \frac{1}{m} \cdot \sqrt{\sum_{k=1}^{\infty} \left[\mathrm{J}_{2\,k - 1}(m\,\pi\,M)\right]^2} 
\end{align}
\paragraph{RMS phasor}
Therefore, the rotating RMS phasor for the emission band of order $m$ is
\begin{align}
\underline{U}_\mathrm{ab}^{\mathrm{B}m}(t) = \left[\frac{2\,U_\mathrm{DC}}{\pi} \cdot \frac{1}{m} \cdot \sqrt{\sum_{k=1}^{\infty} \left[\mathrm{J}_{2\,k - 1}(m\,\pi\,M)\right]^2}~\right] \cdot \e^{\,\j\,2\,m\left[\omega_\mathrm{c}\cdot t + \varphi_\mathrm{c}\right]}.
\end{align}
\paragraph{Harmonics of the amplitude modulation}
\begin{figure}[b!]
	\centering
	\includegraphics[scale=1]{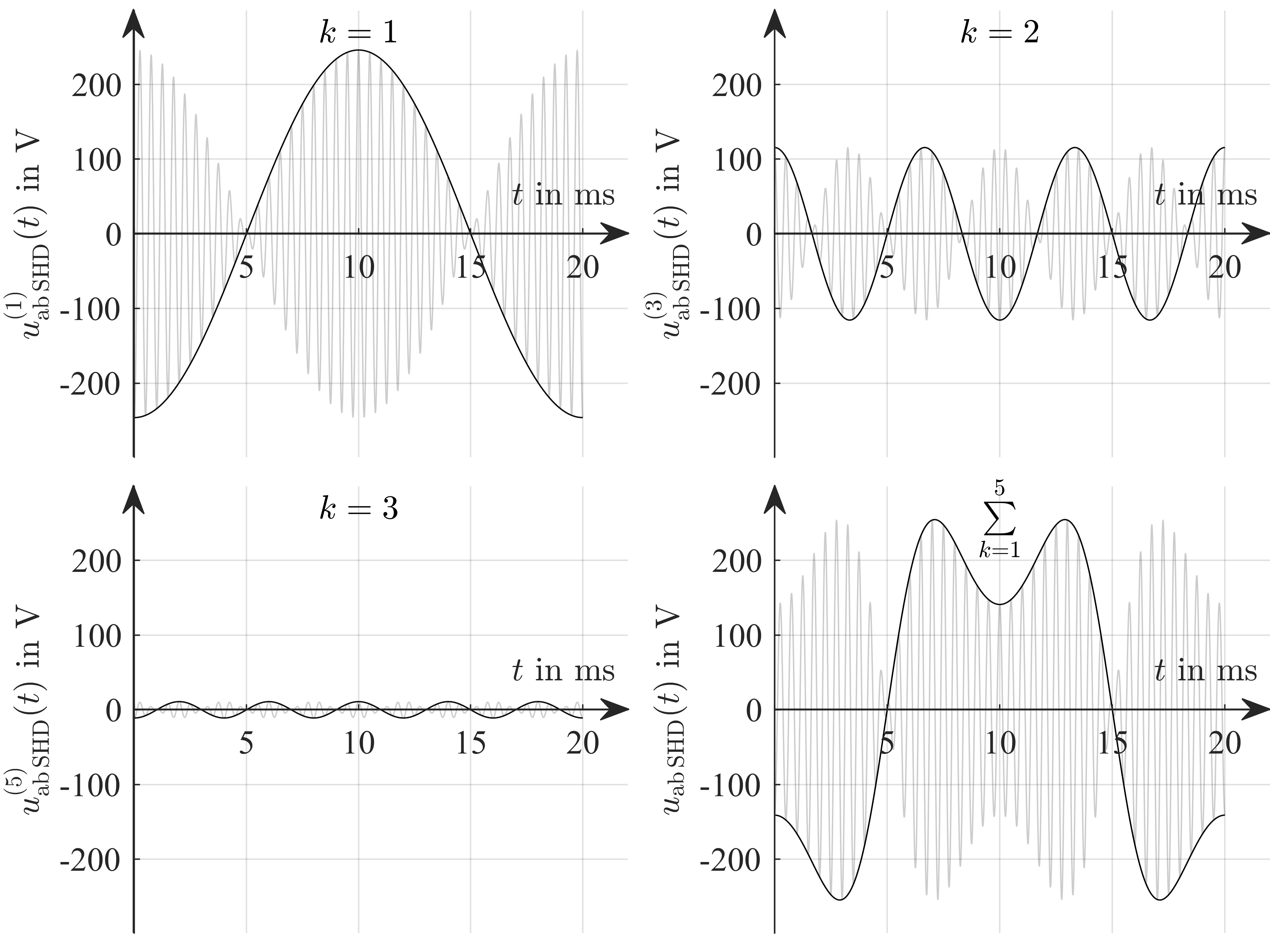}
	\caption{Individual harmonics and their envelopes of the first emission band $m=1$ of the voltage at the terminals of an H-bridge $u_\mathrm{ab\,SHD}(t)$ for natural sampling using a triangular carrier, $f_\mathrm{c} = 1\,\mathrm{kHz}$, $U_\mathrm{DC} = 400\,\mathrm{V}$, $U_\mathrm{ab\,CO} = 230\,\mathrm{V}$, $f_\mathrm{N} = 50\,\mathrm{Hz}$}
	\label{fig:Verlaeufe_Harmonische_erstesEmissionsband}
\end{figure}
The waveform of each emission band contains an infinite number of harmonics of order $\nu=2\,k-1$. The waveform of each harmonic $u_\mathrm{ab\,SHD}^{(\nu)}(m,t) $ can be directly calculated from equation (\ref{eqn:Ergebnis}) by considering only a single component $m$ and $k$ in each sum.
\begin{align}
	u_\mathrm{ab\,SHD}^{(\nu)}(m,t) = \frac{2\,U_\mathrm{DC}}{\pi} \cdot & \frac{1}{m}\cdot \sqrt{2}\cos(2\,m\left[\omega_\mathrm{c}\cdot t + \varphi_\mathrm{c}\right])\cdot \nonumber \\
	&\frac{\mathrm{J}_{\nu}(m\,\pi\,M)}{(-1)^{m+k-1}} \cdot \sqrt{2}\cos(\nu\, \left[\omega_\mathrm{N} \cdot t + \varphi_\mathrm{N}\right]) \label{eqn:Emissionsbandharmonische}
\end{align}
The envelope of each harmonic ${\overline{u}}_\mathrm{ab\,SHD}^{(\nu)}$ of order $\nu=2\,k-1$ of the $m^\mathrm{th}$ emission band can be calculated by taking the absolute value of each component of the amplitude modulation of the rotating pseudo-phasor. This is equivalent to removing the modulation with the emission band frequency $\cos(2\,m\left[\omega_\mathrm{c}\cdot t + \varphi_\mathrm{c}\right])$ from equation (\ref{eqn:Emissionsbandharmonische}).
\begin{align}
	{\overline{u}}_\mathrm{ab\,SHD}^{(\nu)}(m,t) &= \sqrt{2} \cdot \frac{2\,U_\mathrm{DC}}{\pi} \cdot \frac{1}{m} \cdot \frac{\mathrm{J}_{2\,k - 1}(m\,\pi\,M)}{(-1)^{m+k-1}} \cdot \sqrt{2}\cos(\left[2\,k - 1\right] \left[\omega_\mathrm{N} \cdot t + \varphi_\mathrm{N}\right])
\end{align}
In Figure~\ref{fig:Verlaeufe_Harmonische_erstesEmissionsband} the waveforms and envelopes for the first three harmonics ($k=1\ldots3$) and the sum of the first five harmonics of the first emission band $m=1$ are shown for an examplary case.

\FloatBarrier